\documentclass[conference,10pt]{IEEEtran}
\IEEEoverridecommandlockouts
\usepackage[font=footnotesize,skip=0pt]{caption}

\usepackage{cite}
\usepackage{amsmath,amssymb,amsfonts}
\usepackage{algorithmic}
\usepackage{tabularx}
\usepackage{adjustbox}
\usepackage{multirow}
\usepackage{makecell}
\usepackage{graphicx}
\usepackage{textcomp}
\usepackage{comment}
\usepackage{xcolor}
\usepackage{cleveref}
\usepackage{braket}
\usepackage{enumitem}
\usepackage{tikz}
\usepackage{setspace}
\usepackage{stfloats}
\def\BibTeX{{\rm B\kern-.05em{\sc i\kern-.025em b}\kern-.08em
    T\kern-.1667em\lower.7ex\hbox{E}\kern-.125emX}}

\usepackage[compact]{titlesec}

\titlespacing\section{0pt}{0.3\baselineskip}{0.2\baselineskip}
\titlespacing\subsection{0pt}{0.2\baselineskip}{0.1\baselineskip}
\titlespacing\subsubsection{0pt}{0.15\baselineskip}{0.1\baselineskip}

\begin{document}

\title{Next-Generation Quantum Neural Networks: Enhancing Efficiency, Security, and Privacy
}

\author{\IEEEauthorblockN{Nouhaila Innan\IEEEauthorrefmark{1}\IEEEauthorrefmark{2}, Muhammad Kashif\IEEEauthorrefmark{1}\IEEEauthorrefmark{2}, Alberto Marchisio\IEEEauthorrefmark{1}\IEEEauthorrefmark{2}, Mohamed Bennai\IEEEauthorrefmark{3}, Muhammad Shafique\IEEEauthorrefmark{1}\IEEEauthorrefmark{2}}

\IEEEauthorblockA{\IEEEauthorrefmark{1} \small eBrain Lab, Division of Engineering, New York University Abu Dhabi, PO Box 129188, Abu Dhabi, UAE}
\IEEEauthorblockA{\IEEEauthorrefmark{2} \small Center for Quantum and Topological Systems, NYUAD Research
Institute, New York University Abu Dhabi, UAE}
\IEEEauthorblockA{\IEEEauthorrefmark{3} \small Quantum Physics and Spintronics Team, LPMC, Faculty of Sciences Ben M'sick, Hassan II University of Casablanca, Morocco}

{ \small Emails: \{nouhaila.innan, muhammadkashif, alberto.marchisio, muhammad.shafique\}@nyu.edu, mohamed.bennai@univh2c.ma}

}

\IEEEoverridecommandlockouts 

\maketitle

\IEEEpubidadjcol

\begin{abstract}
This paper provides an integrated perspective on addressing key challenges in developing reliable and secure Quantum Neural Networks (QNNs) in the Noisy Intermediate-Scale Quantum (NISQ) era. In this paper, we present an integrated framework that leverages and combines existing approaches to enhance QNN efficiency, security, and privacy. Specifically, established optimization strategies, including efficient parameter initialization, residual quantum circuit connections, and systematic quantum architecture exploration, are integrated to mitigate issues such as barren plateaus and error propagation. Moreover, the methodology incorporates current defensive mechanisms against adversarial attacks. Finally, Quantum Federated Learning (QFL) is adopted within this framework to facilitate privacy-preserving collaborative training across distributed quantum systems. Collectively, this synthesized approach seeks to enhance the robustness and real-world applicability of QNNs, laying the foundation for reliable quantum-enhanced machine learning applications in finance, healthcare, and cybersecurity.

\end{abstract}

\begin{IEEEkeywords}
Quantum Neural Networks, Quantum Machine Learning, Quantum Federated Learning
\end{IEEEkeywords}

\section{Introduction}

Quantum Neural Networks (QNNs) have emerged as a promising paradigm at the intersection of quantum computing and machine learning, offering potential advantages in processing complex data structures and solving computationally intensive tasks~\cite{zaman2023survey}. In the current Noisy Intermediate-Scale Quantum (NISQ) era, characterized by quantum processors with limited qubit counts and susceptibility to noise~\cite{preskill2018quantum}, hybrid quantum-classical QNNs are envisioned to leverage quantum mechanics to achieve high performance in specific applications~\cite{zaman2024comparative, kashif2024computational,kashif2021design}.

Despite their theoretical potential, the practical deployment of QNNs faces several significant challenges inherent to NISQ devices. One of the critical issues is the occurrence of barren plateaus in the optimization landscape, where gradients vanish exponentially with system size~\cite{McClean:2018}, which hinders the effective training of variational quantum circuits. Additionally, the limited number of qubits and their short coherence times constrain the scalability of QNNs, making it challenging to handle high-dimensional data. The presence of noise and decoherence further complicates the reliable execution of quantum circuits, affecting the expressibility and robustness of QNN models~\cite{Kashif_2024_investigating}. Moreover, concerns regarding the security and privacy of quantum machine learning systems, especially in adversarial settings, remain largely unexplored.

To address these challenges, we propose a comprehensive cross-layer methodology aimed at enhancing the efficiency, security, and privacy of QNNs in the NISQ era. Our contributions are as follows:

\begin{itemize}[leftmargin=*]
    \item \textbf{Trainability and Scalability Optimization:} We deploy techniques to mitigate barren plateaus, such as layer-wise training and parameter initialization strategies, and explore quantum circuit cutting methods to enable scalable QNN architectures on limited qubit devices.
    \item \textbf{Noise-Aware Design and Architecture Exploration:} We conduct an in-depth analysis of QNN robustness against various noise models, including phase flip, bit flip, and depolarizing channels, and guide the design of robust QNNs.
    \item \textbf{Security and Privacy Enhancements:} We investigate the vulnerability of QNNs to adversarial attacks and their respective defense mechanisms. Furthermore, we explore the integration of federated learning and encryption techniques to ensure data privacy in distributed QNN scenarios.
    \item \textbf{Application Outlook:} We provide an overview of emerging applications for next-generation QNNs, highlighting their potential impact in fields such as intelligent transportation, finance, and healthcare.
\end{itemize}

\section{Methodology}
\begin{figure*}[ht]
    \centering
    \includegraphics[width=1.0\linewidth]{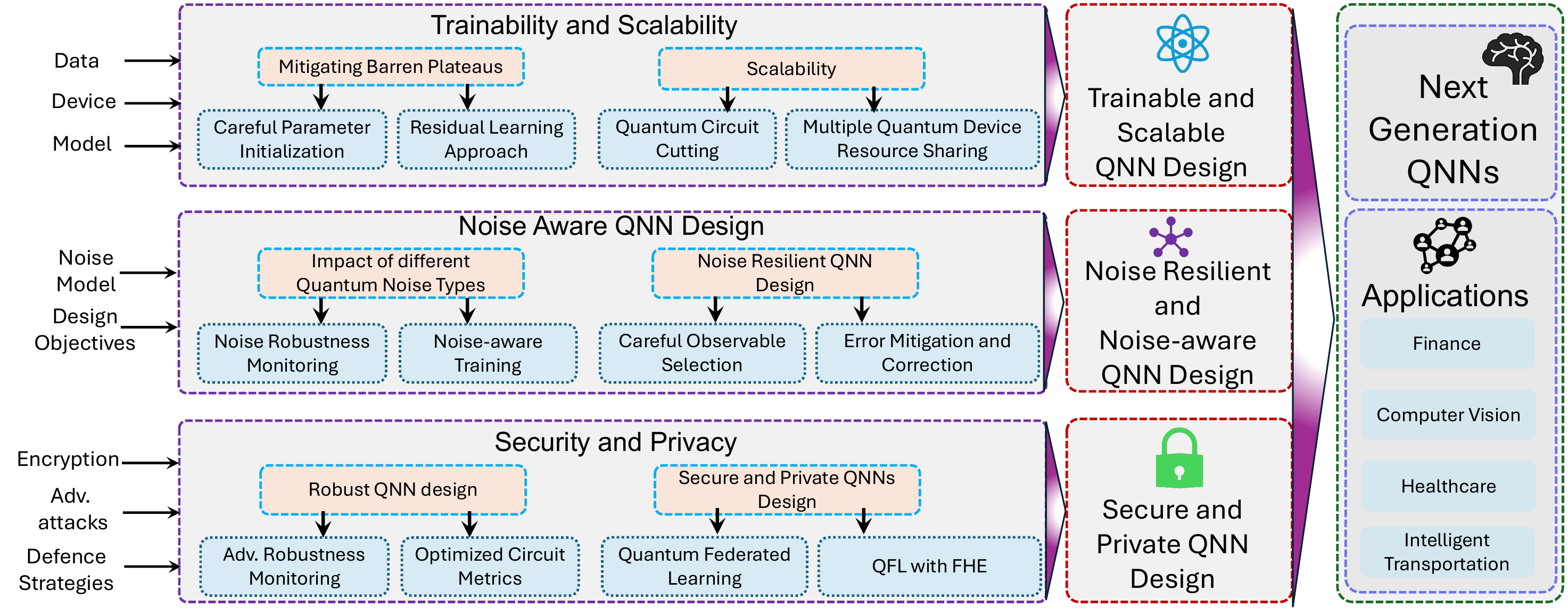}
    \caption{Overview of our methodology to design efficient, robust, and secure QNNs.}
    \label{fig:methoddology_IOLTS}
\end{figure*}

We present an end-to-end framework designed to support the development of next-generation QNNs, with a focus on enhancing their trainability, scalability, and ensuring security and privacy. An overview of our proposed methodology is illustrated in Fig.~\ref{fig:methoddology_IOLTS}.

\subsection{Trainability and Scalability}
\begin{figure}
    \centering
    \includegraphics[width=1.0\linewidth]{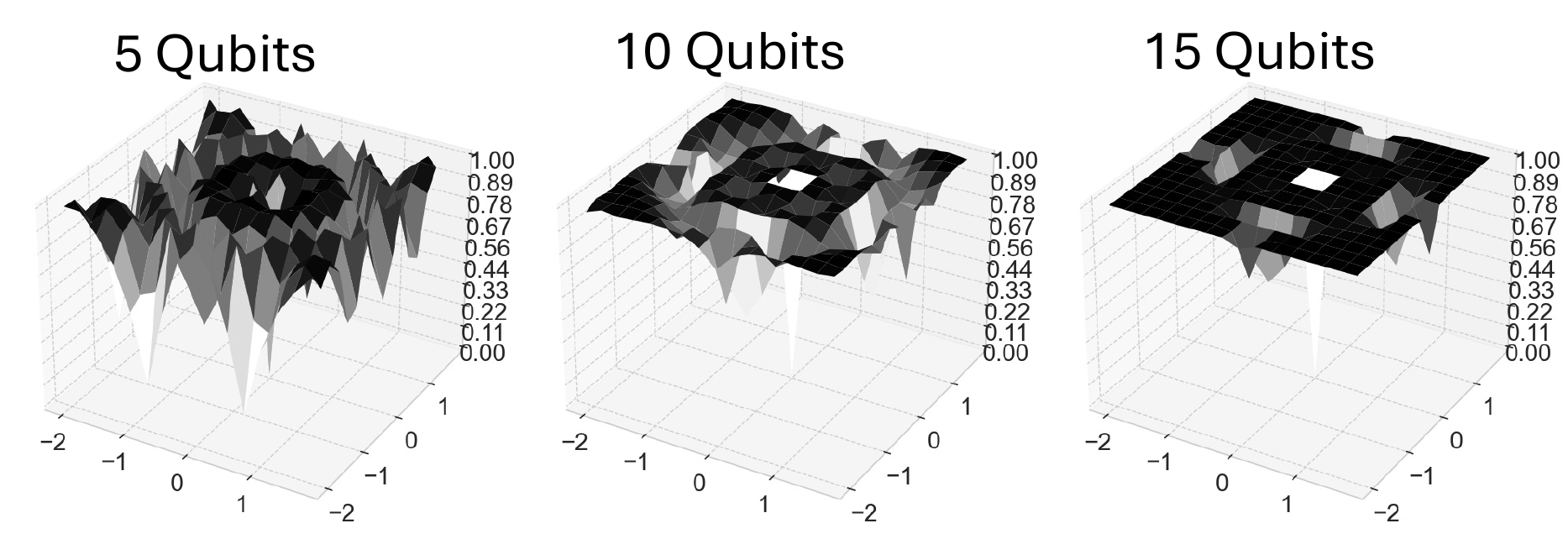}
    \caption{Demonstration of Barren Plateaus.}
    \label{fig:BP_demo_IOLTS}
\end{figure}

\textbf{\textit{Trainability}} is a key challenge in developing QNNs, primarily due to barren plateaus (BP), regions in the optimization landscape where the gradient variance vanishes exponentially with system size~\cite{McClean:2018}, rendering gradient-based training infeasible (see Fig.~\ref{fig:BP_demo_IOLTS}).

\begin{figure}
    \centering
    \includegraphics[width=1.0\linewidth]{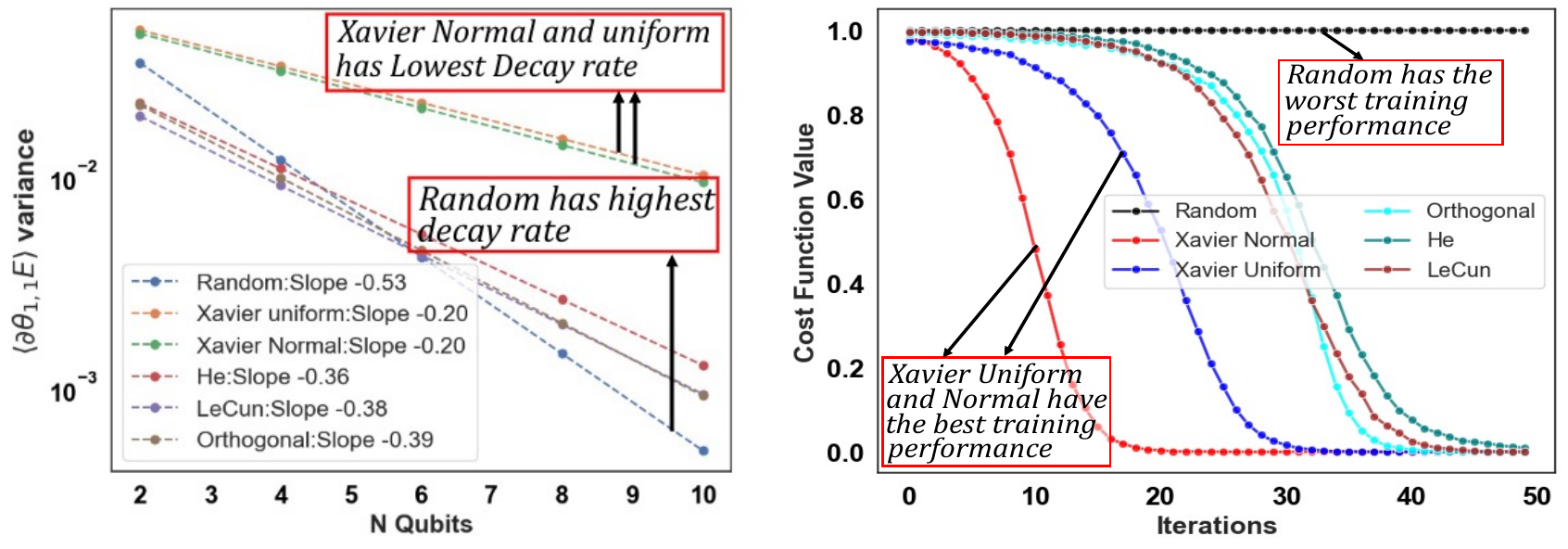}
    \caption{Variance decay and training results with different initialization techniques~\cite{kashif:2024_alleviating}.}
    \label{fig:mit_BP_DATE}
\end{figure}

BPs are linked to random initialization~\cite{kashif:2024_alleviating}, global cost functions~\cite{kashif2023impact}, hardware noise~\cite{kashif:2024_hqnet}, and the expressibility of parameterized quantum circuits (PQCs)~\cite{Kashif:2023_unified}. Mitigation strategies include careful parameter initialization, residual learning, and noise-assisted training. In particular, Xavier initialization has shown superior performance in reducing BP effects~\cite{kashif:2024_alleviating}, as illustrated in Fig.~\ref{fig:mit_BP_DATE}, while narrower initialization ranges further enhance trainability~\cite{kashif:2024_dilemma}. Residual connections, inspired by classical deep networks, preserve gradient flow and enable deeper QNNs~\cite{kashif:2024_resqnets,kashif:2024_resqunns}.


\begin{figure}[h]
    \centering
    \includegraphics[width=1.0\linewidth]{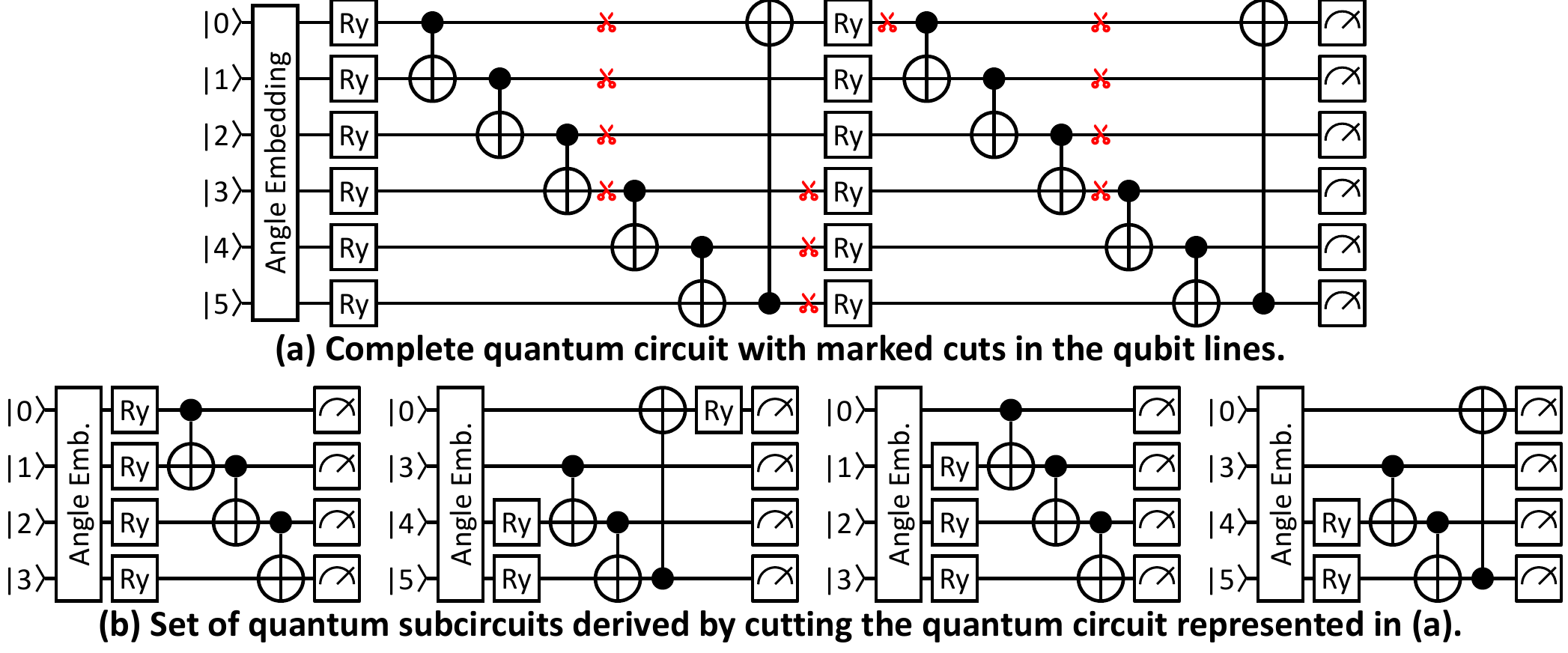}
    \caption{Quantum circuit representation when cutting a 6-qubit circuit into four 4-qubit subcircuits~\cite{marchisio2024cutting}.}
    \label{fig:cutting_example_6to4_IOLTS}
\end{figure}

\textbf{\textit{Scalability}} is constrained by limited qubit counts and noise in NISQ hardware. Current QNN architectures are restricted to small-scale models that fit within the available quantum resources. To overcome this, quantum circuit cutting has been proposed~\cite{marchisio2024cutting}, allowing large circuits to be decomposed into smaller subcircuits. As shown in Fig.~\ref{fig:cutting_example_6to4_IOLTS}, a 6-qubit circuit can be executed using 4-qubit subcircuits. Intermediate measurement results are stored and reused in subsequent subcircuits. Importantly, this approach preserves model accuracy while enabling deployment on limited hardware.

\subsection{Noise-Aware Design and QNN Architecture Exploration}

In the NISQ era, integrating hardware noise considerations into QNN design is vital for practical deployment. Noise-aware training and architecture exploration are essential to identify robust configurations that perform reliably under realistic noise conditions.

\begin{figure}
    \centering
    \includegraphics[width=1.0\linewidth]{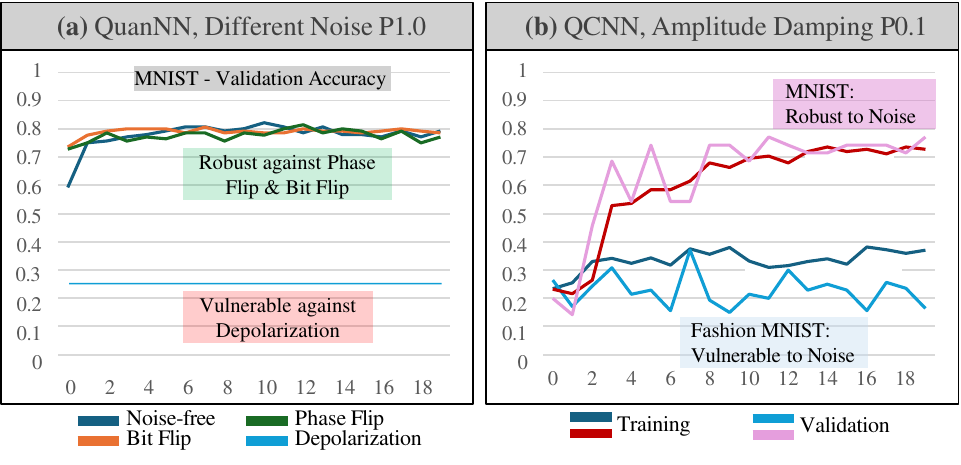}
    \caption{Noise robustness of QNNs under different noise types and tasks. (a) QuanNN's noise robustness under different noise channels with probability 1.0, for the MNIST dataset. (b) QCNN's noise robustness under Amplitude Damping noise with probability 0.1, for the MNIST and Fashion MNIST datasets~\cite{ahmed2025NoisyHQNNs}.}
    \label{fig:NoisyHQNNs_results}
\end{figure}

The work in~\cite{ahmed2025NoisyHQNNs} presents a detailed study on how different quantum noise types affect QNN architectures, specifically Quanvolutional Neural Networks (QuanNNs) and Quantum Convolutional Neural Networks (QCNNs), using real-world datasets. As shown in Fig.~\ref{fig:NoisyHQNNs_results}(a), QuanNNs demonstrate robustness against phase and bit flip noise but degrade under depolarizing noise. In contrast, Fig.~\ref{fig:NoisyHQNNs_results}(b) shows that QCNNs maintain high accuracy on MNIST under amplitude damping noise but struggle with the more complex Fashion-MNIST, illustrating dataset-dependent resilience. This analysis underscores the importance of selecting noise-resilient architectures and informs the design of QNNs better suited to noisy environments. It also highlights when quantum error mitigation or correction becomes crucial to maintain model reliability.


In \cite{kashif:2024_nrqnn}, the impact of quantum noise on QNN trainability is investigated, revealing that BPs arise more readily in noisy settings. The study shows that selecting suitable measurement observables, particularly a custom Hermitian observable aligned with the learning objective, can substantially enhance QNN resilience and trainability across various noise types.


\subsection{Security and Privacy}

After achieving trainability, scalability, and noise resilience, ensuring the security and privacy of QNNs is crucial. Without this, even the most robust architectures remain vulnerable to adversarial threats and data leakage, a critical aspect in domains like healthcare, finance, and cybersecurity.


\subsubsection{Embedding Adversarial Robustness in QNN Design}

Neural networks, including hybrid quantum-classical models, are susceptible to adversarial inputs. Thus, adversarial robustness should be embedded from the design phase. This includes analyzing circuit-level properties such as expressibility, entanglement, and gate configurations. Circuits with high expressibility and controlled entanglement, particularly those using Z-axis controlled rotations, demonstrate improved resistance to adversarial attacks~\cite{10646505, el2024robqunns, maouaki2024designing}.

\begin{figure}
    \centering
    \includegraphics[width=1\linewidth]{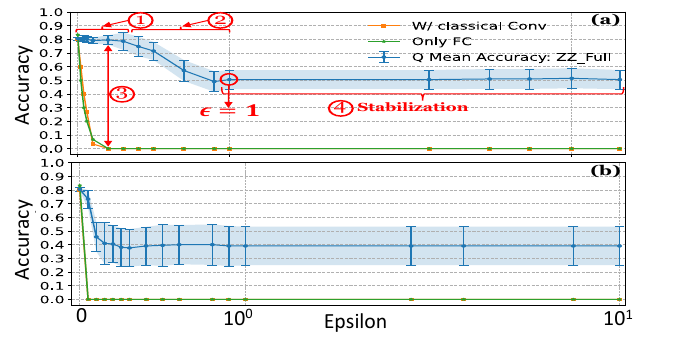}
    \vspace{-0.6cm}
    \caption{Robustness evaluation of classical and quantum models using the ZZ full ansatz against adversarial attacks on the MNIST dataset: (a) FGSM and (b) PGD, tested across varying perturbation strengths~\cite{10646505}.}
    \label{security}
\end{figure}

Rather than relying on post-training defenses, a preemptive adversarial testing phase, simulating attacks like FGSM and PGD, should guide architectural choices. As shown in~\cite{el2025designing}, this helps identify circuits that generalize well even under perturbations. Empirical results show that such circuits achieve up to 60\% greater robustness on datasets like MNIST and Fashion-MNIST at low perturbation levels (see Fig.~\ref{security}).


\subsubsection{Ensuring Privacy through Quantum Federated Learning and Encryption}

QNNs deployed in sensitive domains require decentralized, privacy-preserving training. We adopt a federated learning paradigm, where QNNs are trained locally and only encrypted model updates, not raw data, are shared.

\begin{figure}
    \centering
    \includegraphics[width=1\linewidth]{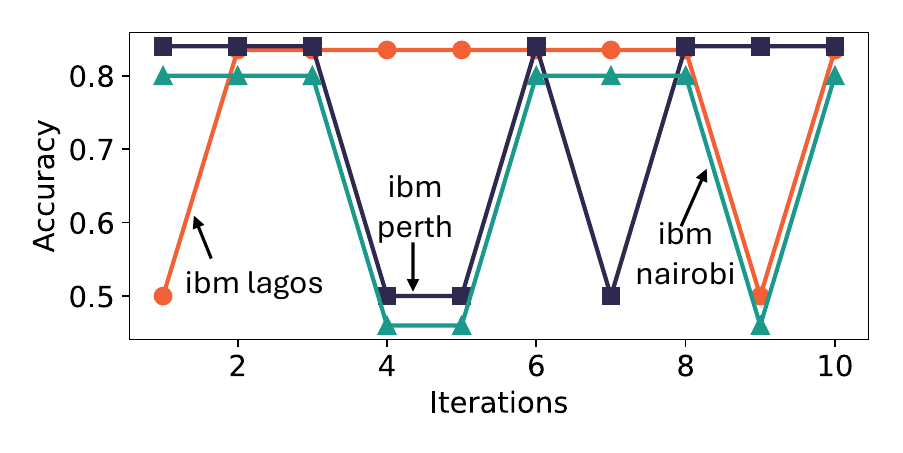}
    \vspace{-0.8cm}
    \caption{Execution of QNNs on real quantum hardware within a federated learning setup, demonstrating stable performance across different IBM Quantum backends~\cite{10651123}.}
    \label{qpu}
\end{figure}

Quantum Federated Learning (QFL)~\cite{10651123} enables distributed quantum nodes to train locally while maintaining data privacy. Each client uses quantum hardware or simulators to train QNNs and transmits only quantum-encoded parameters for global aggregation (see Fig.~\ref{qpu}). This setup is ideal for data-sensitive sectors, such as healthcare and finance.


To further protect updates, we integrate Fully Homomorphic Encryption (FHE)~\cite{dutta2024mqfl}, enabling encrypted aggregation without exposing plaintext parameters. Though FHE can reduce performance, quantum-enhanced models, especially on multimodal data, help retain accuracy. Across diverse datasets, this QFL-FHE pipeline maintains over 70\% test accuracy while preserving privacy (see Table~\ref{qfl}).


\begin{table}[!ht]
\centering
\caption{QFL with FHE experiment results for various datasets. 
}
\resizebox{\linewidth}{!}{%
\begin{tabular}{|c|c|c|c|c|}
\hline
 \textbf{Dataset} & \textbf{Test Loss} & \textbf{Train Accuracy} & \textbf{Test Accuracy} & \textbf{Time (sec)} \\ \hline

 CIFAR-10             & $0.0937$             & $97.90\%$           & $71.12\%$           & $9747.32\pm2.23$ \\ \cline{1-5} 
    DNA Sequence         & $0.782$              & $100.00\%$          & $94.32\%$           & $7123.91\pm2.91$ \\ \cline{1-5} 
    MRI Scan             & $0.360$               & $100.00\%$          & $88.75\%$           & $7851.86\pm3.54$ \\ \cline{1-5} 
     PCOS                 & $1.090$               & $100.00\%$          & $70.15\%$          & $3942.60\pm1.65$ \\ \cline{1-5} 
     RAVDESS                 & $0.83$              & $94.53\%$           & $76.43\%$           & $1140.76\pm1.69$ \\ \cline{1-5} 
     DNA+MRI    & DNA: $0.174$  & DNA: $99.64\%$      & DNA: $95.31\%$      &  \\ 
       Multimodal                   & MRI: $0.713$   & MRI: $100\%$      & MRI: $87.26\%$      & $10314.34\pm6.28$ \\ \hline
\end{tabular}%
}
\label{qfl}
\end{table}

Together, these methods form a robust and secure QNN development pipeline. With adversarial defense and encrypted federated learning, QNNs are now capable of trustworthy deployment in sensitive real-world applications.


\section{Applications}

QNNs have shown significant promise across various domains, demonstrating practical advantages in tasks that demand high accuracy, security, and computational efficiency. Following the development of scalable, noise-aware, and secure QNN architectures, recent research has begun translating these models into real-world applications.
In the financial sector, QNNs have been successfully deployed for fraud detection and loan eligibility prediction \cite{innan2024financial, alami2024comparative}. Privacy-preserving frameworks that integrate federated learning with quantum layers have achieved precision rates exceeding 95\%, even under noisy and distributed settings \cite{innan2024qfnn}. QNNs trained on structured financial data have also achieved up to 98\% accuracy in predicting loan approvals, aided by dropout mechanisms and robust quantum circuit design \cite{innan2024lep}.
In quantum information processing, QNN-based models have enhanced the efficiency of quantum state tomography by reducing the number of required measurements without compromising reconstruction fidelity. These advances are particularly beneficial for scaling to larger quantum systems \cite{innan2024quantum}. While in healthcare, QNNs have enabled precise multi-omics integration for lung cancer classification, uncovering key biomarkers with exceptional diagnostic accuracy \cite{saggi2024mqml}.
In intelligent transportation systems, QNNs have been used to process large-scale traffic data, achieving classification accuracies above 97\% and demonstrating strong robustness under noise, highlighting their potential for deployment in urban mobility infrastructures \cite{innan2025qnn}. Similarly, hybrid quantum-classical models with attention mechanisms have been employed for image super-resolution, offering competitive quality while reducing parameter counts, thus aligning with current hardware limitations in the NISQ era \cite{dutta2025quiet}.
These applications underscore the versatility and practical relevance of QNNs, reinforcing their potential as a foundational technology across sectors where performance, privacy, and resilience are paramount.

\section{Conclusion and Outlook}

In this paper, we presented an integrated, cross-layer methodology for advancing QNNs in the NISQ era. By synthesizing optimization techniques, architecture exploration, noise-aware design, adversarial robustness, and federated privacy-preserving learning, our framework systematically addresses the core limitations of QNNs related to trainability, scalability, and trustworthiness. We demonstrated how approaches such as parameter initialization, residual connections, quantum circuit cutting, and noise-informed design choices can significantly improve model performance on current quantum hardware. Furthermore, we established a foundation for secure and privacy-aware QNN training via adversarial testing and QFL augmented with encryption techniques.

Looking forward, as quantum hardware continues to evolve and mature, our approach offers a scalable path toward deploying robust and secure quantum machine learning models in real-world applications. Future work will involve deeper integration of quantum error correction techniques, exploration of hardware-efficient QNN architectures, and automated quantum architecture search to further streamline development. Additionally, expanding federated and privacy-enhanced quantum learning to edge and cloud environments will be key to enabling widespread adoption. Ultimately, this work serves as a step toward realizing the next-generation QNN systems capable of transforming critical sectors such as healthcare, finance, and intelligent infrastructure.

\section*{Acknowledgment}
This work was supported in part by the NYUAD Center for Quantum and Topological Systems (CQTS), funded by Tamkeen under the NYUAD Research Institute grant CG008, and the Center for Cyber Security (CCS), funded by Tamkeen under the NYUAD Research Institute Award G1104.

\bibliographystyle{ieeetr}
\bibliography{main}

\end{document}